\documentclass[12pt]{iopart}
\usepackage{graphicx}

\begin{document}

\title{Gaussian Enveloped Decoherence of the Atomic States in Quantum Cavity}
\author{G Zhang and Z Song}

\begin{abstract}
We revisit the decoherence of the atomic state in the resonant
Jaynes-Cummings model with the field initially being in a coherent state. We
show that the purity of the atom exhibits oscillating Gaussian dependence on
the time with a width independent of the initial atomic state. It is also
shown that when the atom and the coherent state match each other in phase,
the atomic decoherence is Gaussian time dependence.
\end{abstract}

\pacs{
42.50.Pq, 42.50.Ct, 42.50.Ex  }

\address{School
of Physics, Nankai University, Tianjin 300071, China} %
\ead{songtc@nankai.edu.cn}
\maketitle

\section{Introduction}

Quantum decoherence is at the heart of both the foundations and applications
of quantum physics. Cavity quantum electrodynamics (QED) systems, operating
in the strong coupling regime, have proven to be excellent for the studies
of the entangled atom-photon state \cite{Berman 94}. The experiment
involving the Rydberg atoms in a high Q microwave cavity have opened the way
to the studies of the decoherence dynamics in a mesoscopic system \cite%
{Brune 96}. Theoretically, the simplest model that captures the physics of
such a hybrid system is the Jaynes--Cummings (JC) model \cite{Jaynes
63,Cummings 65}. It is one of the few exactly solvable models in quantum
optics and predicts several interesting effects such as the vacuum field
Rabi oscillations \cite{Stenholm 73,Meystre 75,Meystre 82}, collapses and
revivals of Rabi oscillations in the coherent field \cite{Yoo 85}.
Remarkably, it was noticed that the atom is to a good approximation in a
pure state in the middle between the collapse and revival \cite%
{Gea-Banacloche 90,Gea-Banacloche 91}.

On the other hand, being related to the quantum measurement theory and the
quantum decoherence problems, the influence induced by the spin bath on the
decoherence dynamics of a central system have also attracted much attention
\cite{Quan 06}. It was shown that the decoherence induced by coupling a
system with an environment may display universal features: the decay of
quantum coherences in the system is Gaussian for the specific initial
environment state \cite{Cucchietti 05,Zurek 07,Cucchietti 07}

In this paper, we revisit the decoherence of the atomic state in the
resonant JC model with the field initially being in a coherent state and
elaborate the dynamic evolution of the purity. Closed analytical expressions
for the purity of the central 2-level atoms are obtained. We observe that
the similar behavior as that of the spin bath occurs in such a hybrid
system. Both the first collapse and the amplitude of subsequent oscillation
exhibit a Gaussian decay behavior.

The paper is organized as follows. In Sec. II, we introduce the main
properties of JC model. In Sec. III, we evaluate the purity of the central
2-level atom. Section IV presents our summary and conclusion. In the
Appendix we derive the main formula needed to evaluate various sums used in
the text.

\section{The JC model}

Starting point of the analysis is the JC model, which consists of a single
atom coupled to a single mode cavity. The two possible states of the atom
are the ground state $\left\vert g\right\rangle $, and its excited state $%
\left\vert e\right\rangle $. The model and the subject we discussed are the
same as the one previously\ studied by Gea--Banacloche \cite{Gea-Banacloche
90,Gea-Banacloche 91}, who showed that the atom is to a good approximation
in a pure state in the middle between the collapse time $t_{c}$ and revival
time$\ t_{r}$. In the following, we will not only reproduce the results of
\cite{Gea-Banacloche 90,Gea-Banacloche 91} but also show that this
remarkable phenomenon is a part of the dynamical process for a special case.
The model Hamiltonian at the resonance has the form

\begin{eqnarray}
H &=&\lambda \left( \sigma _{+}a+\sigma _{-}a^{\dag }\right) +\frac{1}{2}%
\omega _{a}\sigma _{z}+\omega a^{\dagger }a  \label{H_JC} \\
\sigma _{+} &=&\left( \sigma _{-}\right) ^{\dagger }=\left\vert
e\right\rangle \left\langle g\right\vert ,\sigma _{z}=\left\vert
e\right\rangle \left\langle e\right\vert -\left\vert g\right\rangle
\left\langle g\right\vert  \nonumber
\end{eqnarray}%
where $a^{\dag }$\ is the creation operators of photon with frequency $%
\omega $, $\omega _{a}$ is the atomic transition frequency, and $\lambda $
is the cavity--atom coupling constant. The aim here is to study the dynamics
of a given initial state. A general initial state of the system has the form%
\begin{equation}
\left\vert \Psi \left( 0\right) \right\rangle =\left( C_{g}\left\vert
g\right\rangle +C_{e}\left\vert e\right\rangle \right) {\sum_{n=0}^{\infty }}%
C_{n}\left\vert n\right\rangle  \label{Initial}
\end{equation}%
\ where $|C_{e}|^{2}+|C_{g}|^{2}=1$. Of central importance is the excitation
number
\begin{equation}
\mathcal{N}=\frac{1}{2}\sigma _{z}+a^{\dagger }a+\frac{1}{2}
\label{Exciton N}
\end{equation}%
is a conserved quantity, i.e., $\left[ \mathcal{N},H\right] =0$, which makes
it easy to diagonalize the Hamiltonian, since the atom-field eigenspaces are
only two-dimensional. It also makes the dynamics of states involving several
subspaces simple. Nevertheless, the dynamics of states that have many
significant energy-state components can show considerable complexity.

Introducing a unitary transformation

\begin{equation}
\mathcal{R}\left( \theta ,\phi \right) =\mathrm{e}^{\mathrm{i}\left( \theta
\sigma _{z}/2+\phi a^{\dagger }a\right) }  \label{R}
\end{equation}%
which generate the phases $\theta $ and $\phi $ on the atomic and cavity
states
\begin{equation}
\mathcal{R}\left( \theta ,\phi \right) \left( C_{g}\left\vert g\right\rangle
+C_{e}\left\vert e\right\rangle \right) \left\vert n\right\rangle =\mathrm{e}%
^{\mathrm{i}n\phi }\left( \mathrm{e}^{-\mathrm{i}\theta /2}C_{g}\left\vert
g\right\rangle +\mathrm{e}^{\mathrm{i}\theta /2}C_{e}\left\vert
e\right\rangle \right) \left\vert n\right\rangle ,
\end{equation}%
\ we have $\mathcal{R}H\mathcal{R}^{\dagger }=\tilde{H}$, where

\begin{eqnarray}
\tilde{H} &=&\lambda \left( \sigma _{+}\tilde{a}+\sigma _{-}\tilde{a}^{\dag
}\right) +\frac{1}{2}\omega _{a}\sigma _{z}+\omega \tilde{a}^{\dag }\tilde{a}
\label{H_t} \\
&=&\lambda \left( \tilde{\sigma}_{+}a+\tilde{\sigma}_{-}a^{\dag }\right) +%
\frac{1}{2}\omega _{a}\tilde{\sigma}_{z}+\omega a^{\dagger }a  \nonumber
\end{eqnarray}%
with $\tilde{a}=\mathrm{e}^{-\mathrm{i}\left( \phi -\theta \right) }a$, $%
\tilde{\sigma}_{+}=\mathrm{e}^{-\mathrm{i}\left( \phi -\theta \right)
}\sigma _{+}$ and $\tilde{\sigma}_{z}=\sigma _{z}$. It indicates that
Hamiltonians $H$ and $\tilde{H}$ share the same eigenfunctions by
transformation of the basis $\left\{ \left\vert n\right\rangle \right\}
\rightarrow \left\{ \mathrm{e}^{\mathrm{i}n\phi }\left\vert n\right\rangle
\right\} $ or $\left\vert g\right\rangle \rightarrow \mathrm{e}^{-\mathrm{i}%
\theta /2}\left\vert g\right\rangle $ and $\left\vert e\right\rangle
\rightarrow \mathrm{e}^{\mathrm{i}\theta /2}\left\vert e\right\rangle $.
Remarkably, in the case of $\theta =\phi $, we have $\mathcal{R}H\mathcal{R}%
^{\dagger }=\tilde{H}=H$, which shows the invariance of the Hamiltonian
under the transformation $\mathcal{R}\left( \theta ,\theta \right) $. We
will show that, when dealing with the coherent cavity state, this feature
leads to an interesting and important phenomenon. We will demonstrate the
strong dependence of the dynamics of the atomic purity on the relative phase
of the atom and the cavity field.

We shall only consider the resonant case of $\omega =\omega _{a}$. Then at
time $t$, state $\left\vert \Psi \left( 0\right) \right\rangle $ evolves to

\begin{eqnarray}
\left\vert \Psi \left( t\right) \right\rangle =\sum_{n=1}^{\infty }\left[
C_{n}C_{g}\cos \left( \sqrt{n}\lambda t\right) -\mathrm{i}C_{n-1}C_{e}\sin
\left( \sqrt{n}\lambda t\right) \right] \left\vert g,n\right\rangle &&
\label{Psi_t} \\
+\sum_{n=1}^{\infty }\left[ C_{n}C_{e}\cos \left( \sqrt{n+1}\lambda t\right)
-\mathrm{i}C_{n+1}C_{g}\sin \left( \sqrt{n+1}\lambda t\right) \right]
\left\vert e,n\right\rangle ], &&  \nonumber
\end{eqnarray}

We concern the reduced density matrix of the atom, which has the form%
\begin{equation}
\rho _{A}\left( t\right) =\left(
\begin{array}{cc}
a & b \\
b^{\ast } & 1-a%
\end{array}%
\right)  \label{RDM_A}
\end{equation}%
where%
\begin{eqnarray}
a &=&\sum_{n=1}^{\infty }\left[ \left\vert C_{n}C_{g}\right\vert ^{2}\cos
^{2}\left( \sqrt{n}\lambda t\right) +\left\vert C_{n-1}C_{e}\right\vert
^{2}\sin ^{2}\left( \sqrt{n}\lambda t\right) \right.  \label{ab} \\
&&\left. -2\mathrm{Im}\left( C_{n}C_{n-1}^{\ast }C_{g}C_{e}^{\ast }\right)
\cos \left( \sqrt{n}\lambda t\right) \sin \left( \sqrt{n}\lambda t\right) %
\right] ,  \nonumber \\
b &=&\sum_{n=1}^{\infty }\left[ \left\vert C_{n}\right\vert
^{2}C_{g}C_{e}^{\ast }\cos \left( \sqrt{n}\lambda t\right) \cos \left( \sqrt{%
n+1}\lambda t\right) \right.  \nonumber \\
&&\left. +C_{n-1}C_{n+1}^{\ast }C_{e}C_{g}^{\ast }\sin \left( \sqrt{n}%
\lambda t\right) \sin \left( \sqrt{n+1}\lambda t\right) \right.  \nonumber \\
&&\left. +\mathrm{i}C_{n}C_{n+1}^{\ast }\left\vert C_{g}\right\vert ^{2}\cos
\left( \sqrt{n}\lambda t\right) \sin \left( \sqrt{n+1}\lambda t\right)
\right.  \nonumber \\
&&\left. -\mathrm{i}C_{n-1}C_{n}^{\ast }\left\vert C_{e}\right\vert ^{2}\sin
\left( \sqrt{n}\lambda t\right) \cos \left( \sqrt{n+1}\lambda t\right) %
\right] .  \nonumber
\end{eqnarray}%
As a measure of the degree of coherence, the purity the atom can be
expressed as%
\begin{equation}
P(t)=\mathrm{Tr}\left( \rho _{A}^{2}\right) =a^{2}+\left( 1-a\right)
^{2}+2\left\vert b\right\vert ^{2},  \label{purity}
\end{equation}%
where Tr$\left( ...\right) $\ denotes the trace on the cavity field.

\section{Decoherence of a two-level atom}

With the time evolution of the reduced density matrix of the atom, we can
investigate the dynamical behavior of the atom, which has been employed to
calculate the inversion and the purity for the case of initial coherent
state. The initial state has the form

\begin{equation}
\left\vert \Psi \left( 0\right) \right\rangle =\left( C_{g}\left\vert
g\right\rangle +C_{e}\left\vert e\right\rangle \right) \left\vert \alpha
\right\rangle  \label{coherent}
\end{equation}%
where%
\begin{equation}
\left\vert \alpha \right\rangle =\mathrm{e}^{-\left\vert \alpha \right\vert
^{2}/2}\sum_{n=0}^{\infty }\frac{\alpha ^{n}}{\sqrt{n!}}\left\vert
n\right\rangle ,\alpha =\mathrm{e}^{-\mathrm{i}\phi }\sqrt{{\bar{n}}}.
\label{alpha}
\end{equation}%
It is has been found that the initial Rabi-oscillations concerning the
probability of being in a given atomic state decay on a timescale called the
collapse time, $t_{c}=2/\lambda $, but then revive after a much longer time,%
\textbf{\ }$t_{r}=2\pi \sqrt{\bar{n}}/\lambda $\textbf{\ }\cite{Barnett
97,Eberly 80}\textbf{.}

Here we discuss the time dependence of the atomic purity in a long time
scale. For $\left\vert \alpha \right\vert ^{2}$ (or $\bar{n}$) $\gg 1$, we
have%
\begin{equation}
C_{n}=\mathrm{e}^{-\left\vert \alpha \right\vert ^{2}/2}\frac{\alpha ^{n}}{%
\sqrt{n!}}=\frac{\alpha }{\sqrt{n}}C_{n-1}\approx \mathrm{e}^{-\mathrm{i}%
\phi }C_{n-1}.  \label{flat}
\end{equation}%
Remarks on flat condition. In the following we take $\phi =0$\ for the sake
of simplicity, since factor $\mathrm{e}^{-\mathrm{i}\phi }$\ can be mapped
on the atomic state by $\left( C_{g},C_{e}\right) \rightarrow \left( \mathrm{%
e}^{-\mathrm{i}\phi /2}C_{g}+\mathrm{e}^{\mathrm{i}\phi /2}C_{e}\right) $
according to our previous analysis. Then in the following derivation, we
simply consider the coefficients $C_{g}$\ and $C_{e}$\ as complex numbers.
On the other hand, for sufficiently large value of $\bar{n}$, the Poisson
distribution is an approximation to the normal (or Gaussian) distribution,
i.e.,%
\begin{equation}
C_{n}C_{n-1}\approx C_{n}^{2}=\exp \left( -\bar{n}\right) \frac{\bar{n}^{n}}{%
n!}\approx \frac{1}{\sqrt{2\pi \bar{n}}}\exp \left[ -\frac{\left( n-\bar{n}%
\right) ^{2}}{2\bar{n}}\right] .  \label{Poisson Normal}
\end{equation}%
On the other hand, the Poissonian function peaks sharply around $\bar{n}$.
Then for a nontrivial function $F(\sqrt{n},1/\sqrt{n})$, one can take the
approximation%
\begin{equation}
C_{n}^{2}F(\sqrt{n},1/\sqrt{n})\approx \frac{1}{\sqrt{2\pi \bar{n}}}\exp %
\left[ -\frac{\left( n-\bar{n}\right) ^{2}}{2\bar{n}}\right] F(\frac{\sqrt{%
\bar{n}}}{2}+\frac{n}{2\sqrt{\bar{n}}},\frac{3}{2\sqrt{\bar{n}}}-\frac{n}{2%
\sqrt{\bar{n}^{3}}}).  \label{distribution}
\end{equation}%
Furthermore, in the limit of $\bar{n}\gg 1$\ the summation over $n$ can be
done exactly by virtue of Euler--Maclaurin formula. In the Appendix we
derive the main formula needed to evaluate various sums used in the text.
Accordingly, we obtain the analytic form for reduced density matrix elements
are%
\begin{eqnarray}
a &\approx &\frac{1}{2}+\left[ \frac{1}{2}\left( \left\vert C_{g}\right\vert
^{2}-\left\vert C_{e}\right\vert ^{2}\right) \cos \left( \frac{4\sqrt{\bar{n}%
}t}{t_{c}}\right) \right. \\
&&\left. +\left\vert C_{g}C_{e}\right\vert \sin \theta \sin \left( \frac{4%
\sqrt{\bar{n}}t}{t_{c}}\right) \right] \exp \left( -\frac{2t^{2}}{t_{c}^{2}}%
\right) ,  \nonumber \\
b &\approx &\left[ \left\vert C_{g}C_{e}\right\vert \cos \theta \cos \left(
\frac{\pi t}{t_{r}}\right) +\mathrm{i}\frac{1}{2}\sin \left( \frac{\pi t}{%
t_{r}}\right) \right] \exp \left( -\frac{\pi ^{2}}{8\bar{n}}\frac{t^{2}}{%
t_{r}^{2}}\right)  \nonumber \\
&&+\mathrm{i}\left[ \frac{1}{2}\left( \left\vert C_{g}\right\vert
^{2}-\left\vert C_{e}\right\vert ^{2}\right) \sin \left( \frac{4\sqrt{\bar{n}%
}t}{t_{c}}\right) \right.  \nonumber \\
&&\left. +\left\vert C_{g}C_{e}\right\vert \sin \theta \cos \left( \frac{4%
\sqrt{\bar{n}}t}{t_{c}}\right) \right] \exp \left( -\frac{2t^{2}}{t_{c}^{2}}%
\right)  \nonumber
\end{eqnarray}

From equation (\ref{purity}), we have%
\begin{eqnarray}
P(t) &=&\frac{1}{2}+2\left[ \left\vert C_{g}C_{e}\right\vert ^{2}\cos
^{2}\theta \cos ^{2}\left( \frac{\pi t}{t_{r}}\right) \right. \left. +\frac{1%
}{4}\sin ^{2}\left( \frac{\pi t}{t_{r}}\right) \right] \exp \left( -\frac{%
\pi ^{2}}{4\bar{n}}\frac{t^{2}}{t_{r}^{2}}\right)  \label{P_t1} \\
&&+\frac{1}{2}\left[ \left( \left\vert C_{g}\right\vert ^{2}-\left\vert
C_{e}\right\vert ^{2}\right) ^{2}+4\left\vert C_{g}C_{e}\right\vert ^{2}\sin
^{2}\theta \right] \exp \left( -\frac{4t^{2}}{t_{c}^{2}}\right)  \nonumber \\
&&+\sin \left( \frac{\pi t}{t_{r}}\right) \left[ \left( \left\vert
C_{g}\right\vert ^{2}-\left\vert C_{e}\right\vert ^{2}\right) \sin \left(
\frac{4\sqrt{\bar{n}}t}{t_{c}}\right) \right.  \nonumber \\
&&\left. +2\left\vert C_{g}\right\vert \left\vert C_{e}\right\vert \sin
\theta \cos \left( \frac{4\sqrt{\bar{n}}t}{t_{c}}\right) \right] \exp \left(
-\frac{\pi ^{2}}{8\bar{n}}\frac{t^{2}}{t_{r}^{2}}\right) \exp \left( -\frac{%
2t^{2}}{t_{c}^{2}}\right) .  \nonumber
\end{eqnarray}%
\begin{figure}[tbp]
\begin{center}
\includegraphics[ bb=15 200 560 650, width=18 cm, clip]{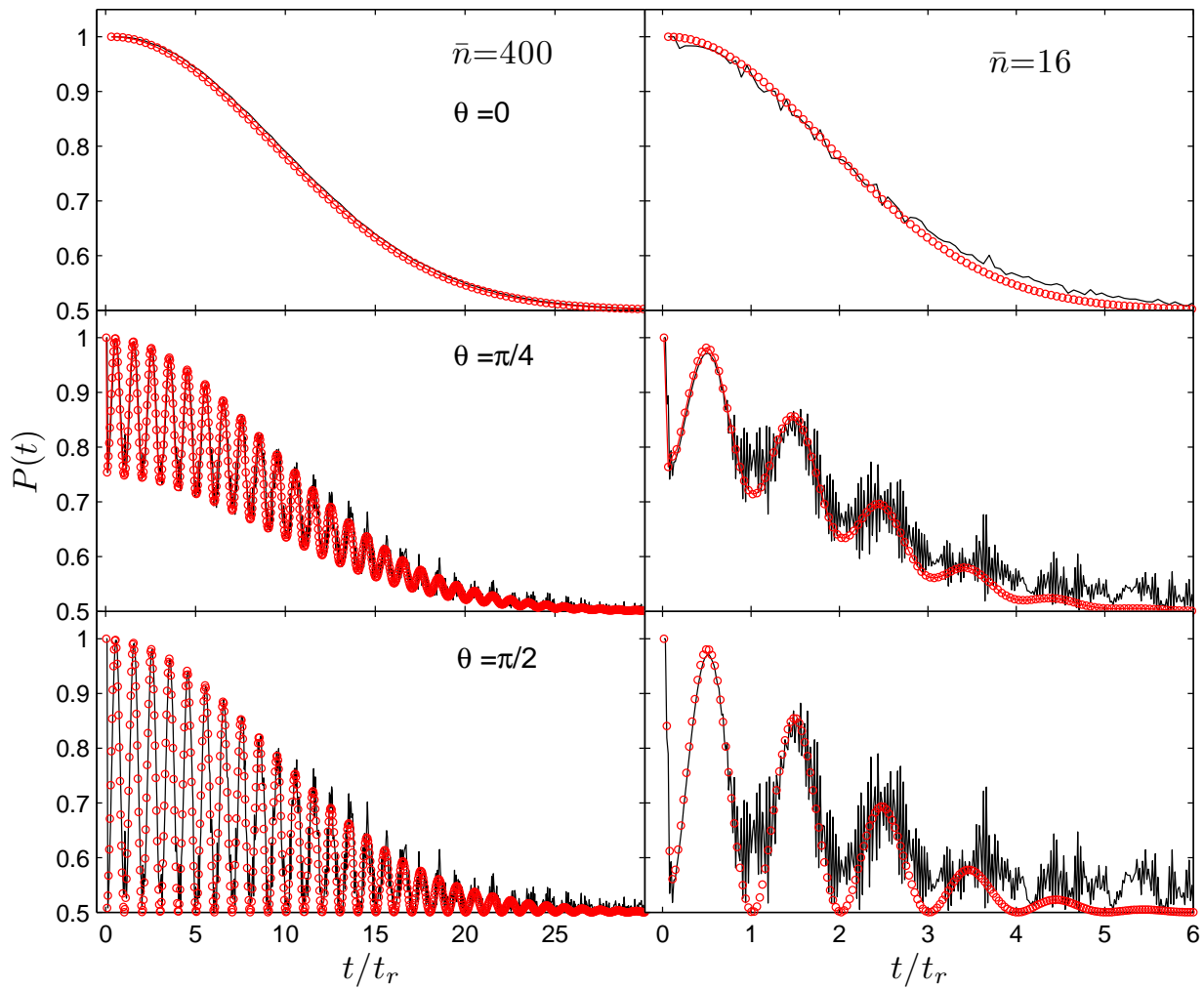}
\end{center}
\caption{(Color online) Plots of the purity obtained by the exact
diagonalization (solid line) and the approximate analytical expression
equation (\protect\ref{P_t1}) (empty circle) at various relative phases $%
\protect\theta $, for the initial atomic state with $\left\vert
C_{g}\right\vert =\left\vert C_{e}\right\vert $\ and the initial coherent
state of photons with average number $\bar{n}=400$ and $\bar{n}=16$.}
\label{figure1}
\end{figure}

\begin{figure}[tbp]
\begin{center}
\includegraphics[ bb=66 300 530 525, width=16 cm, clip]{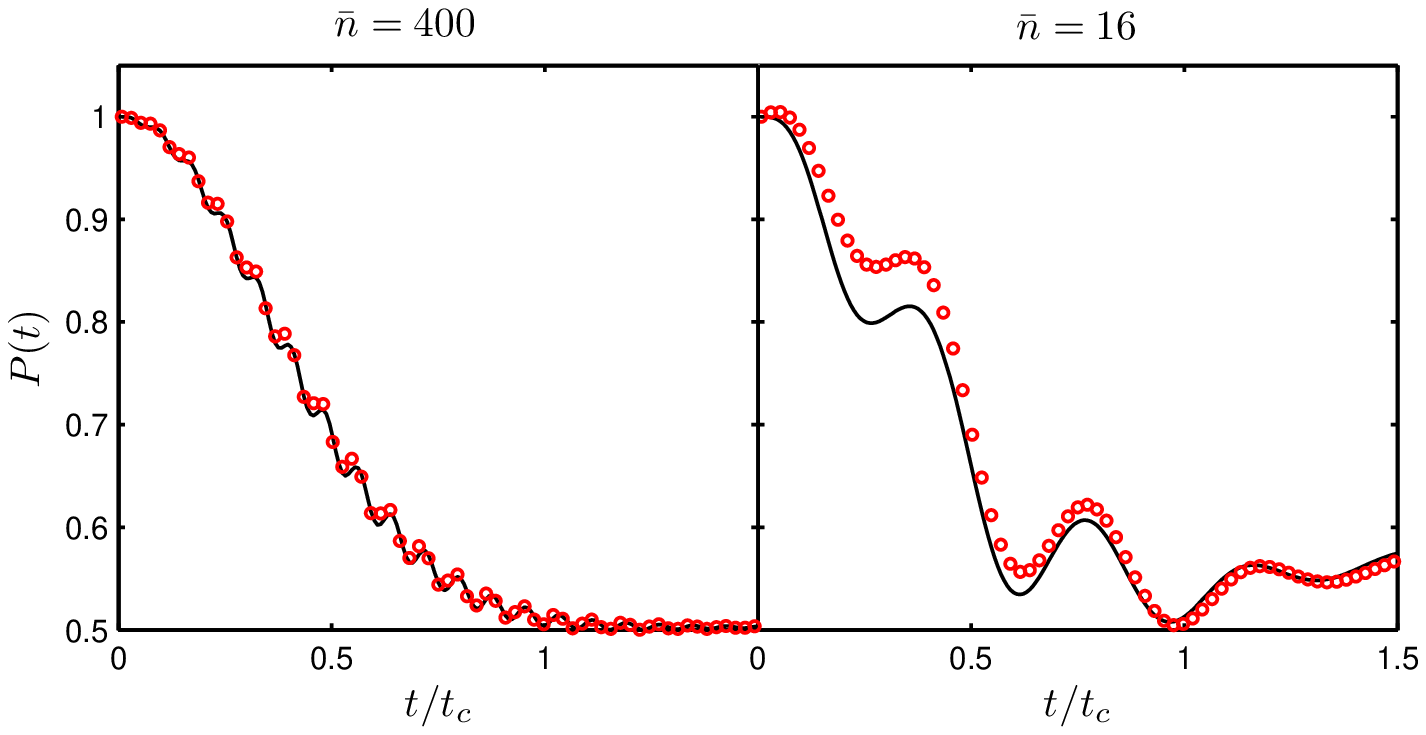}
\end{center}
\caption{(Color online) The same as figure \protect\ref{figure1} but in the
small time scale and the case of $\protect\theta =\protect\pi /2$.}
\label{figure2}
\end{figure}
We can see that, at small time region $t\ll t_{r}$, the term containing $%
\exp \left( -4t^{2}/t_{c}^{2}\right) $ is a transient state and is dominant
at the beginning of the evolution. This Gaussian decay behavior is
independent of the mean photon number $\bar{n}$\textbf{.} After the
transient relaxation, the purity of the atom exhibits oscillating Gaussian
dependence on the time. The width of Gaussian function is independent of the
values of $C_{g}$ and $C_{e}$. The initial atomic state determines the
amplitude of the oscillation. Both the first collapse and the amplitude of
subsequent oscillation exhibit a Gaussian decay behavior.\ Using the
analytic expressions for the purity amplitude equation (\ref{P_t1}), we
estimate the period of the oscillation to be the same as that of the Rabi
oscillation.\ Obviously, purity dynamics depends on the parameter of the
system as well as the initial state. In this work we show that the relative
phase between the initial atom and cavity field has far more important
influence on the purity dynamics.\ Let us consider two interesting special
cases: $\left\vert C_{g}C_{e}\right\vert =0$\ and $\left\vert
C_{g}\right\vert =\left\vert C_{e}\right\vert $. In first case,\ $C_{g}=0$ (
or $C_{e}=0$), we have
\begin{eqnarray}
P_{\max }(t) &=&\frac{1}{2}+\frac{1}{2}\sin ^{2}\left( \frac{\pi t}{t_{r}}%
\right) \exp \left( -\frac{\pi ^{2}}{4\bar{n}}\frac{t^{2}}{t_{r}^{2}}\right)
+\frac{1}{2}\exp \left( -\frac{4t^{2}}{t_{c}^{2}}\right)   \label{P_max} \\
&&\pm \sin \left( \frac{\pi t}{t_{r}}\right) \sin \left( \frac{4\sqrt{\bar{n}%
}t}{t_{c}}\right) \exp \left( -\frac{\pi ^{2}}{8\bar{n}}\frac{t^{2}}{%
t_{r}^{2}}\right) \exp \left( -\frac{2t^{2}}{t_{c}^{2}}\right) ,  \nonumber
\end{eqnarray}%
i.e., the amplitude of the oscillations becomes maximum. In second case, $%
C_{g}=\mathrm{e}^{\mathrm{i}\theta }C_{e}$, we have%
\begin{eqnarray}
P_{\theta }(t) &=&\frac{1}{2}+\frac{1}{2}\left[ 1-\sin ^{2}\theta \cos
^{2}\left( \frac{\pi t}{t_{r}}\right) \right] \exp \left( -\frac{\pi ^{2}}{4%
\bar{n}}\frac{t^{2}}{t_{r}^{2}}\right)  \\
&&+\frac{1}{2}\sin ^{2}\theta \exp \left( -\frac{4t^{2}}{t_{c}^{2}}\right)
\nonumber  \label{P_theta} \\
&&+\sin \left( \frac{\pi t}{t_{r}}\right) \sin \theta \cos \left( \frac{4%
\sqrt{\bar{n}}t}{t_{c}}\right) \exp \left( -\frac{\pi ^{2}}{8\bar{n}}\frac{%
t^{2}}{t_{r}^{2}}\right) \exp \left( -\frac{2t^{2}}{t_{c}^{2}}\right) .
\nonumber
\end{eqnarray}%
It shows that after the transient process, the amplitude of the oscillations
only depends on the relative phase $\theta $. We also note that $P_{\theta
}(t)$ behaves as the purity for various values of the coefficients $C_{g}=%
\mathrm{e}^{-\mathrm{i}\delta /2}\left\vert C_{g}\right\vert $ and $C_{e}=%
\mathrm{e}^{\mathrm{i}\delta /2}\left\vert C_{e}\right\vert $ by simply
replacing $\sin ^{2}\theta $ in the equation (\ref{P_theta})$\ $with $%
1-4\left\vert C_{g}C_{e}\cos \delta \right\vert ^{2}$.\ Thus in the
following numerical simulations, we only demonstrate the case of $\left\vert
C_{g}\right\vert =\left\vert C_{e}\right\vert $ for simplicity.

We note that in the case of $\theta =0$%
\begin{equation}
P_{\min }(t)=\frac{1}{2}+\frac{1}{2}\exp \left( -\frac{\pi ^{2}}{4\bar{n}}%
\frac{t^{2}}{t_{r}^{2}}\right) ,  \label{P_min}
\end{equation}%
which corresponds to the envelope of the pattern $P(t)$\ for arbitrary
initial atomic state.

The above analysis shows two important characteristics of the decoherence
dynamics. At first, the decoherence occurs dramatically at the very
beginning for an arbitrary initial atomic state except the case of $%
C_{g}=C_{e}$. Secondly, after the transient decoherence, the amplitude of
the oscillating purity strongly depends on the initial phase difference
between the atom and the field. When the initial atom and the field are
in-phase or opposite-phase ($\phi -\theta =0$, $\pi $), the atom has a
relatively long coherent time. When they are orthogonal-phase ($\phi -\theta
=\pi /2$, $3\pi /2$), the atom acquires maximal oscillating amplitude of
decoherence.

In order to verify the above analysis some numerical simulations are
performed. In figure \ref{figure1}, we plot the equation (\ref{purity}) for $%
\bar{n}=400$\ and $16$\ cases with different values of $\theta $\ and $%
\left\vert C_{g}\right\vert =\left\vert C_{e}\right\vert $. As comparison,
we also plot the equation (\ref{P_t1}) accordingly. We can see that the
analytical results match well with the simulation results, especially in
large $\bar{n}$ case and during the first several periods of oscillation.
figure\textbf{\ }\ref{figure2}\textbf{\ }is the same as the plot in figure%
\textbf{\ }\ref{figure1}\textbf{\ }but for small time scale and\textbf{\ }$%
\theta =\pi /2$ to demonstrate the transient process explicitly.

It is also worthwhile to mention that the Gaussian decay of the decoherence
is the direct result of the coherent state environment. We note that the
expression equation (\ref{P_t1}) is obtained under the two conditions: (i)
distribution function $C_{n}$ is flat as equation (\ref{flat}); (ii) we can
use the approximation equation (\ref{distribution}) near the mean photon
number $\bar{n}$. The result for\textbf{\ }$\theta =0$\textbf{\ }may promise
important potential applications in quantum-information processing since
Gaussian time dependence of the decoherence factor would suggest a different
more frequent error correction than the exponential dependence.

\begin{figure}[tbp]
\begin{center}
\includegraphics[ bb=58 246 580 556, width=16 cm, clip]{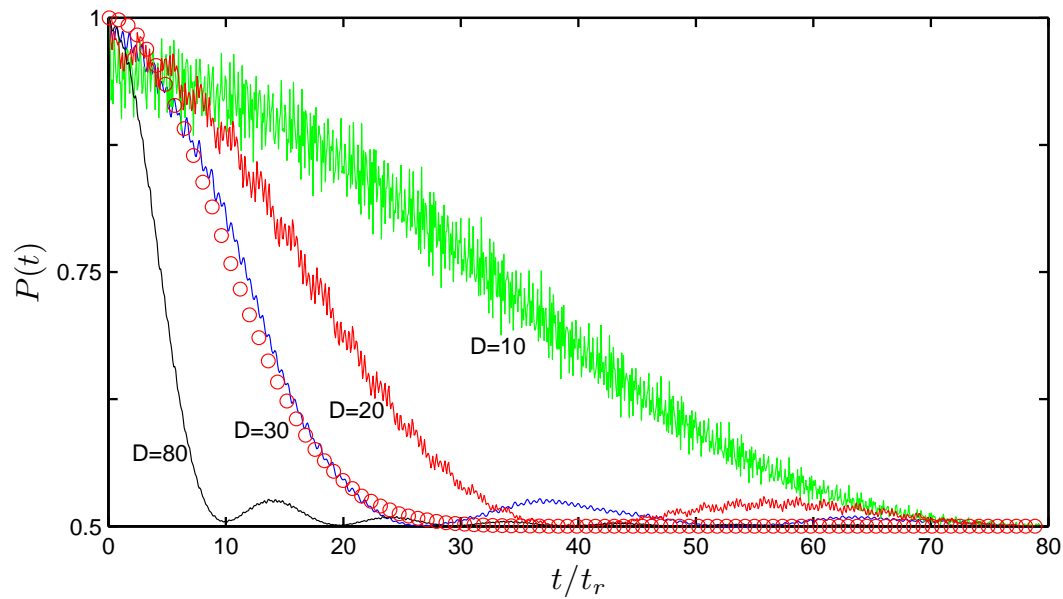}
\end{center}
\caption{(Color online) Plots of the purity for the top-hat initial field
states with average number $\bar{n}=400,$\ $D=10$, $20$, $30$, and $80$, and
the approximate analytical expression (empty circle) for the initial
coherent state of the field with the same average number. It indicates that
the atomic coherence decay differs from the Gaussian function except the
special case.}
\label{figure3}
\end{figure}

\section{Summary}

In conclusion, considering a system consisting of a two-level atom,
initially prepared in a coherent superposition of two levels, interacting
with a coherent state of the field, we show that the dynamics of the atomic
purity are sensitive to the relative phase between the atom and the cavity
field. We also observe that the purity of the atom exhibits oscillating
Gaussian dependence on the time with a width independent of the initial
atomic state. Our results may have a great potential for future applications
in quantum optical device.

\section*{Acknowledgment}

We acknowledge the support of the CNSF (Grant Nos. 10874091 and
2006CB921205).

\appendix{}

This appendix contains the formulas needed to evaluate various sums used in
the paper.

\textit{Sum 1. }Consider the sum

\begin{equation}
S_{1}=\frac{1}{\sqrt{2\pi \bar{n}}}\sum_{n=1}^{\infty }\exp \left[ -\frac{%
\left( n-\bar{n}\right) ^{2}}{2\bar{n}}+\mathrm{i}\frac{\lambda t}{2\sqrt{n}}%
\right] .
\end{equation}%
According to equation (\ref{distribution}), we have%
\begin{equation}
S_{1}\approx \frac{1}{\sqrt{2\pi \bar{n}}}\sum_{n=1}^{\infty }\exp \left[ -%
\frac{\left( n-\bar{n}\right) ^{2}}{2\bar{n}}+\mathrm{i}\frac{\lambda t}{4%
\sqrt{\bar{n}}}\left( 3-\frac{n}{\bar{n}}\right) \right] .
\end{equation}%
From Euler--Maclaurin formula, we replace the sum by the integral%
\begin{eqnarray}
S_{1} &\approx &\frac{1}{\sqrt{2\pi \bar{n}}}{\int_{infty}^{\infty }}\exp %
\left[ -\frac{\left( x-\bar{n}\right) ^{2}}{2\bar{n}}+\mathrm{i}\frac{%
\lambda t}{4\sqrt{\bar{n}}}\left( 3-\frac{x}{\bar{n}}\right) \right] \mathrm{%
d}x \\
&=&\exp \left( -\frac{\lambda ^{2}t^{2}}{32\bar{n}^{2}}+\mathrm{i}\frac{%
\lambda t}{2\sqrt{\bar{n}}}\right) ,  \nonumber
\end{eqnarray}%
where the Gaussian integral formula

\begin{equation}
\frac{1}{\sigma \sqrt{2\pi }}{\int_{-\infty }^{\infty }}\exp \left[ -\frac{%
\left( \beta -\overline{\beta }\right) ^{2}}{2\sigma ^{2}}-\mathrm{i}\beta t%
\right] \mathrm{d}\beta =\exp \left( -\frac{\sigma ^{2}t^{2}}{2}-\mathrm{i}%
\overline{\beta }t\right)  \label{G}
\end{equation}%
has been used.

\textit{Sum 2, 3. }Taking the similar procedure one can calculate the
following two sums

\begin{equation}
S_{2}=\frac{1}{\sqrt{2\pi \bar{n}}}\sum_{n=1}^{\infty }\exp \left[ -\frac{%
\left( n-\bar{n}\right) ^{2}}{2\bar{n}}\right] \cos ^{2}\left( \sqrt{n}%
\lambda t\right)
\end{equation}%
and%
\begin{equation}
S_{3}=\frac{1}{\sqrt{2\pi \bar{n}}}\sum_{n=1}^{\infty }\exp \left[ -\frac{%
\left( n-\bar{n}\right) ^{2}}{2\bar{n}}\right] \cos \left( \sqrt{n}\lambda
t\right) \sin \left( \sqrt{n+1}\lambda t\right) .
\end{equation}%
Actually, using the Gaussian integral formulae%
\begin{equation}
\frac{1}{\sigma \sqrt{2\pi }}{\int_{-\infty }^{\infty }}\exp \left[ -\frac{%
\left( \beta -\overline{\beta }\right) ^{2}}{2\sigma ^{2}}\right] \cos
\left( \beta t\right) \mathrm{d}\beta =\cos \left( \overline{\beta }t\right)
\exp \left( -\frac{\sigma ^{2}t^{2}}{2}\right)  \label{G1a}
\end{equation}%
and%
\begin{equation}
\frac{1}{\sigma \sqrt{2\pi }}{\int_{-\infty }^{\infty }}\exp \left[ -\frac{%
\left( \beta -\overline{\beta }\right) ^{2}}{2\sigma ^{2}}\right] \sin
\left( \beta t\right) \mathrm{d}\beta =\sin \left( \overline{\beta }t\right)
\exp \left( -\frac{\sigma ^{2}t^{2}}{2}\right)  \label{G1b}
\end{equation}%
we have%
\begin{eqnarray}
S_{2} &\approx &\frac{1}{\sqrt{2\pi \bar{n}}}{\int_{-\infty }^{\infty }}\exp %
\left[ -\frac{\left( x-\bar{n}\right) ^{2}}{2\bar{n}}\right] \cos ^{2}\left(
\sqrt{x}\lambda t\right) \mathrm{d}x \\
&\approx &\frac{1}{2}+\frac{1}{2\sqrt{2\pi \bar{n}}}{\int_{-\infty }^{\infty
}}\exp \left[ -\frac{\left( x-\bar{n}\right) ^{2}}{2\bar{n}}\right] \cos %
\left[ \left( \sqrt{\bar{n}}+\frac{x}{\sqrt{\bar{n}}}\right) \lambda t\right]
\mathrm{d}x  \nonumber \\
&=&\frac{1}{2}+\exp \left[ -\frac{\lambda ^{2}t^{2}}{2}\right] \cos \left(
\sqrt{\bar{n}}\lambda t\right)  \nonumber
\end{eqnarray}%
and%
\begin{eqnarray}
S_{3} &\approx &\frac{1}{2\sqrt{2\pi \bar{n}}}{\int_{-\infty }^{\infty }}%
\exp \left[ -\frac{\left( x-\bar{n}\right) ^{2}}{2\bar{n}}\right] \left\{
-\sin \left( \frac{x\lambda t}{4\sqrt{\bar{n}^{3}}}-\frac{3\lambda t}{4\sqrt{%
\bar{n}}}\right) \right. \\
&&\left. +\sin \left[ \left( \frac{x}{\sqrt{\bar{n}}}-\frac{x}{4\sqrt{\bar{n}%
^{3}}}\right) \lambda t+\left( \sqrt{\bar{n}}+\frac{3}{4\sqrt{\bar{n}}}%
\right) \lambda t\right] \right\} \mathrm{d}x  \nonumber \\
&=&\frac{1}{2}\exp \left[ -\frac{\bar{n}t^{2}}{2}\left( \frac{\lambda }{%
\sqrt{\bar{n}}}-\frac{\lambda }{4\sqrt{\bar{n}^{3}}}\right) ^{2}\right] \sin
\left( 2\sqrt{\bar{n}}\lambda t+\frac{\lambda t}{2\sqrt{\bar{n}}}\right)
\nonumber \\
&&+\frac{1}{2}\exp \left[ -\frac{\bar{n}t^{2}}{2}\left( \frac{\lambda }{4%
\sqrt{\bar{n}^{3}}}\right) ^{2}\right] \sin \left( \frac{\lambda t}{2\sqrt{%
\bar{n}}}\right)  \nonumber \\
&\approx &\frac{1}{2}\exp \left( -\frac{\lambda ^{2}t^{2}}{2}\right) \sin
\left( 2\sqrt{\bar{n}}\lambda t\right) +\frac{1}{2}\exp \left( -\frac{%
\lambda ^{2}t^{2}}{32\bar{n}^{2}}\right) \sin \left( \frac{\lambda t}{2\sqrt{%
\bar{n}}}\right) .  \nonumber
\end{eqnarray}

\end{document}